\documentclass{iaus}
\usepackage{graphicx}
\usepackage{natbib}

\newcommand{\apjl}{\textit{ApJL}}
\newcommand{\apj}{\textit{ApJ}}
\newcommand{\mnras}{\textit{MNRAS}}
\newcommand{\aap}{\textit{A\&A}}
\newcommand{\aj}{\textit{AJ}}
\newcommand{\araa}{\textit{ARAA}}

\newcommand\yr{{\;\rm yr}}

\newcommand\Msun{{\;\rm\,M_\odot}}
\newcommand\kms{{\;\rm km\; s^{-1}}}
\newcommand\pc{{\;\rm\,pc}}
\newcommand\kpc{{\;\rm kpc}}
\newcommand\simgt{\lower.5ex\hbox{$\; \buildrel > \over \sim \;$}}
\newcommand\simlt{\lower.5ex\hbox{$\; \buildrel < \over \sim \;$}}

\newcommand\siggbc{\Sigma_{\rm GBC}}
\newcommand\sigdiff{\Sigma_{\rm diff}}
\newcommand\sigsfr{\Sigma_{\rm SFR}}
\newcommand\Pth{P_{\rm th}}
\newcommand\tsf{t_{\rm SF}}

\title[Star Formation in Galactic Disks]{Star Formation and Gas Dynamics in Galactic Disks: \\ 
Physical Processes and Numerical Models}
\author[E.~C. Ostriker ]   %% give here short author list %%
{Eve C. Ostriker
}

\affiliation{Department of Astronomy, University of Maryland, College Park, MD 20742; ostriker@astro.umd.edu}

\pubyear{2010}
\volume{270}  %% insert here IAU Symposium No.
\pagerange{XXX}
\jname{Computational Star Formation}
\editors{J. Alves, B. Elmegreen, J. Girart, \&  V. Trimble, eds.}
\begin{document}

\maketitle

\begin{abstract}
Star formation depends on the available gaseous ``fuel'' as well as
galactic environment, with higher specific star formation rates where
gas is predominantly molecular and where stellar (and dark matter)
densities are higher.  The partition of gas into different thermal
components must itself depend on the star formation rate, since a
steady state distribution requires a balance between heating (largely
from stellar UV for the atomic component) and cooling.  In this
presentation, I discuss a simple thermal and dynamical equilibrium
model for the star formation rate in disk galaxies, where the basic
inputs are the total surface density of gas and the volume density of
stars and dark matter, averaged over $\sim \kpc$ scales.  Galactic
environment is important because the vertical gravity of the stars and
dark matter compress gas toward the midplane, helping to establish the
pressure, and hence the cooling rate.  In equilibrium, the star
formation rate must evolve until the gas heating rate is high enough
to balance this cooling rate and maintain the pressure imposed by the local
gravitational field.  In addition to discussing the formulation of
this equilibrium model, I review the current status of numerical
simulations of multiphase disks, focusing on measurements of
quantities that characterize the mean properties of the diffuse ISM.
Based on simulations, turbulence levels in the diffuse ISM appear
relatively insensitive to local disk conditions and energetic driving
rates, consistent with observations.  It remains to be determined,
both from observations and simulations, how mass exchange processes
control the ratio of cold-to-warm gas in the atomic ISM.

%\keywords{Keyword1, keyword2, keyword3, etc.}
\end{abstract}

\firstsection % if your document starts with a section,

\section{Introduction}

Disk galaxies are gas-rich systems, with a multi-phase, highly
structured interstellar medium (ISM). Within the ISM, star formation
takes place in giant molecular clouds (GMCs), sometimes concentrated
in spiral arms. The rate and character of star formation are
influenced by physical processes from sub-pc to multi-kpc scales
\citep{2007ARA&A..45..565M}.  In spite of the complexity of the ISM
and star formation at small scales, there are nevertheless clear
correlations between the large-scale rate at which stars are born, and
the properties of the ISM and (intra-)galactic environment on large
($\sim$ kpc) scales.

As discussed by Frank Bigiel at this meeting (see also
\citealt{Bigiel08}, and references therein), in regions of galaxies
where the gaseous surface density $\Sigma \simlt 100 \Msun \pc^{-2}$, the star
formation rate closely follows the surface density of molecular gas.
This can be understood in terms of having an essentially constant 
star formation timescale, $\tsf\sim 2\times 10^9\yr$, within molecular gas 
(which is observed to be in organized in gravitationally bound clouds with 
properties that are similar in different galaxies).
As a consequence, $\sigsfr \propto \Sigma$ in regions where the
molecular gas dominates the atomic gas.  For regions where atomic gas
dominates (primarily in the outer parts of galaxies), $\sigsfr$
instead varies as a steeper power of $\Sigma$.  In addition to this
superlinear behavior, there is considerable scatter in the relation
between $\sigsfr$ vs. $\Sigma$ at low surface density, suggesting that
one or more other parameters, in addition to $\Sigma$, controls the
star formation rate. 

Indeed, recent examination of the correlation of $\sigsfr$ with
``non-interstellar'' galactic environmental properties has revealed
interesting dependences, indicating that in the outer parts of galaxies,
both the specific star formation rate  and the ratio of
molecular-to-atomic gas increase roughly linearly with the
\textit{stellar} surface density $\Sigma_s$ \citep{Ler08}.
Previously, \citet{BR06} found an approximately linear increase of the
molecular content with the estimated dynamic pressure of the ISM, and
this is evident in the sample analyzed by \citet{Ler08} as well.
The physical reason for the relationship between molecular
content (and star formation) and pressure has not, however, been clear from
these empirical studies.

Observations of star formation pose a number of challenges: Why is
there an increase in the slope of $\sigsfr \propto \Sigma^{1+p}$ in
going from molecular- to atomic-dominated regions? What is the
physical reason for the empirical relation between ISM pressure and
star formation; more generally, how do galactic parameters such as
$\Sigma_s$, the velocity dispersions of stars and of gas, and spiral 
structure affect $\sigsfr$?  Is it possible to
explain the observed behavior of $\sigsfr$ using simplified
theoretical models, and what is required in numerical simulations in
order to reproduce observed star formation relationships?  Recent
theoretical work has taken on these challenges with increasing success; a
key to these advances has been a more sophisticated treatment of both
the ISM and the galactic environment. For example, \citet{KO09a}
found, using numerical simulations of the ISM and a cooling function
allowing multiple phases, star formation rates and proportions between
self-gravitating and diffuse gas similar to the observations of
\citet{BR06} and \citet{Ler08} provided that turbulent driving is
included; for non-turbulent models, the proportion of self-gravitating
gas was found to be much too high.

\section{A thermal/dynamical equilibrium model for $\sigsfr$}

Motivated by recent observations as well as simulations and earlier
theory, \citet{2010ApJ...OML}
(hereafter OML) have developed a simple model for star formation
regulation in multiphase, turbulent ISM disks. In essence, the OML
model combines three basic principles: (1) the diffuse (atomic)
component of the ISM is in approximate thermal equilibrium, with a
density (and pressure) proportional to the heating rate; (2) the
diffuse component of the ISM is in approximate dynamical equilibrium,
with the pressure at any height above the galactic midplane given by
the weight of the overlying gas; (3) UV from young stars provides most
of the heating for the atomic component of the ISM, with star
formation taking place only within the gravitationally-bound component
of the ISM.  These principles have been individually established and
extensively studied (over several decades) in the astrophysical
literature.  \citet{1969ApJ...155L.149F} combined (1) and (2) to
conclude that the diffuse atomic gas in the local Milky Way must consist of
a two-phase cloud/intercloud medium.  In this and subsequent
treatments of thermal and dynamical equilibrium, the heating rate 
has generally been treated as an independent (empirical) parameter. But,
by including (3) together with (1) and (2), OML obtained a
closed system representing a local patch of a disk galaxy.  For this
closed system, the partition of gas into phases and the star formation
rate are obtained self-consistently.

In the OML model, the (simplified) ISM is treated as
having two components, one consisting of diffuse gas (including both
high-density cold atomic cloudlets and a low-density warm atomic
intercloud medium), and the other consisting of 
gravitationally-bound clouds (GBCs).  Although hot gas is also present
in the ISM, it is a tiny fraction of the mass, and 
fills $\simlt 20\%$ of the volume \citep{2001ASPC..231..294H} (OML
describe how to correct for this effect). 
For galaxies with normal
metallicity, the GBCs would represent giant molecular clouds,
including their atomic shielding layers.  Averaged over $\sim \kpc$
scales (which may contain many or few individual GBCs), the 
total surface density of the GBC component is $\siggbc$, and the total 
surface density of the diffuse component is $\sigdiff$.

The diffuse component is assumed to be in vertical
dynamical equilibrium (as has been verified by numerical simulations;
e.g. \citealt{PO07,KO09b}), with the vertical gravity (from the diffuse
gas, the GBC component, the stellar disk, and the dark matter halo)
balanced by the difference between midplane and external values of
thermal pressure $\Pth$, turbulent pressure $\rho v_z^2$, and magnetic
stresses $(8\pi)^{-1}(B^2 - 2 B_z^2)$.  Because cooling times 
are short compared to other timescales, the diffuse gas is 
assumed to be in thermal equilibrium, with the additional provision
that both warm and cold phases are present.  
This allows a range of
pressures between $P_{\rm min, cold}$ and $P_{\rm max, warm}$; 
for the model of OML, it is
assumed that the pressure is equal to the geometric mean of these
limits, $P_{\rm two-phase}\equiv (P_{\rm min, cold} P_{\rm max,
  warm})^{1/2}$.   For atomic gas, heating is generally 
dominated by the UV and cooling by collisionally-excited lines 
\citep{Wol03}, which 
yields $P_{\rm two-phase} \propto J_{\rm UV}$.  (Note that other
heating -- e.g. cosmic rays and shocks -- can be more important for
very dense, shielded cores and very hot gas, respectively.)
Finally, the OML model assumes that the rate of star
formation is proportional to the total surface density $\siggbc$ of gas
in the GBC component, $\sigsfr=\siggbc/\tsf=(\Sigma -\sigdiff)/\tsf$.

Vertical dynamical equilibrium within the diffuse layer is expressed as 
$P_{\rm tot} \equiv \alpha P_{\rm th} = \Sigma_{\rm diff} \langle g_z \rangle /2$,
where the mean vertical gravity is 
\begin{equation}
\langle g_z \rangle 
\approx \pi G (\sigdiff + 2\siggbc) +2 (2 G \rho_{\rm sd})^{1/2} \sigma_z;
\label{gz-eq}
\end{equation}
$\rho_{\rm sd}$ is the midplane density of stars plus dark matter,
$\sigma_z$ is the total vertical velocity dispersion of the diffuse
gas, and the total pressure is larger than the thermal pressure by a
factor $\alpha$ (see below).  %Note that in equation (\ref{gz-eq}), 
The GBC component contributes more strongly (per unit mass) to the
gravity because its scale height is smaller 
than that of the diffuse gas.

If $n^2\Lambda(T)$ is the cooling rate
per unit volume and $n\Gamma$ is the heating rate per unit volume,
then the two-phase pressure is given by
\begin{eqnarray}
\frac{P_{\rm two-phase}}{k} 
&\equiv& \left(n_{\rm min, cold} T_{\rm min, cold} n_{\rm max,warm} T_{\rm max, warm}
\right)^{1/2}\cr
&=&\Gamma 
\frac{\left( T_{\rm min, cold}  T_{\rm max, warm}\right)^{1/2} }
{\left[\Lambda(T_{\rm min, cold })\Lambda(T_{\rm max,warm})\right]^{1/2}},
\end{eqnarray}
where we have used the equilibrium condition $\Gamma = n \Lambda$ for both  
phases.
Cooling of the cold atomic medium is dominated by metals 
(in particular, C II) so that $\Lambda \propto
Z_{\rm gas}$, while heating is dominated by the photoelectric effect
with $\Gamma \propto Z_{\rm dust} J_{\rm UV}$; since $T_{\rm min, cold
}$ and $T_{\rm max,warm}$ are relatively independent of the heating
rate \citep{Wol95}, this yields $P_{\rm two-phase} \propto J_{\rm UV}
$ if $Z_{\rm dust}/Z_{\rm gas} = const$.  The terms $Z_{\rm gas}$ and
$Z_{\rm dust}$ represent the ratios of metals and dust to hydrogen,
respectively. The mean UV intensity is
affected by radiative transfer through the diffuse gas, 
but for modest optical depth in the
diffuse gas the relation $J_{\rm UV} \propto \sigsfr$ is expected to
hold.  In addition, a larger fraction of the UV escapes from GBCs 
if $Z_d$ is very sub-Solar, which increases $J_{\rm UV}$ for a
given $\sigsfr$ (this effect is quite uncertain, but might increase 
$J_{\rm UV}$ by a factor $\sim 2$).
In thermal equilibrium with $\Pth \sim P_{\rm two-phase}$, 
the midplane pressure is therefore expected to vary roughly as 
$\Pth \propto \sigsfr$, with a somewhat larger coefficient for very
low-metallicity regions. 

Combining the thermal equilibrium relation $\Pth = P_{\rm th,0}
\sigsfr/\Sigma_{\rm SFR, 0}$ (normalized using the Solar neighborhood 
thermal pressure $P_{\rm th,0}$
and star formation rate $\Sigma_{\rm SFR, 0}$)
with the dynamical equilibrium relation 
$\Pth = \Sigma_{\rm diff} \langle g_z \rangle /(2\alpha)$ and the star 
formation relation $\sigsfr=\siggbc/\tsf$, we obtain 
\begin{eqnarray}
\frac{\siggbc}{\sigdiff}&=&\frac{\langle g_z \rangle}{g_*}
%\cr &\propto& 
\propto \pi G (\sigdiff + 2\siggbc) +2 (2 G \rho_{\rm sd})^{1/2} \sigma_z.
\label{ratio-eq}
\end{eqnarray}
Here, $g_* = 2\alpha P_{\rm th,0}/(\Sigma_{\rm SFR, 0} \tsf)$;
for fiducial parameters, this acceleration is 
$g_* \sim \pc \ {\rm Myr}^{-2}$.  

It is interesting to compare outer and inner disks.  In outer disks
(similar to the Solar neighborhood and beyond, in galaxies like the
Milky-Way), diffuse gas dominates the total so that $\siggbc \ll
\sigdiff \approx \Sigma$; in addition, the term depending on
$\rho_{\rm sd}$ dominates the gravity $g_z$.  In this regime, the
relation $\sigsfr \propto \siggbc \propto \Sigma \sqrt{\rho_{\rm sd}}$
is therefore expected to hold.  Physically, this regime may be thought
of as the result of star formation increasing until the heating it
provides is sufficient to balance cooling at the (dynamically-imposed)
midplane pressure.  If there is too little gas in the GBC component,
the star formation rate would be extremely low, and the UV field would
be very weak. A very low heating rate could not
maintain a warm medium at the pressure imposed by the local gravitational
field, so that (a portion of the) 
warm gas would condense out and become cold clouds.
These cold clouds would collect to create more GBCs, which would then
initiate star formation, raising the local UV radiation field until
heating balances cooling.  Given the low gravity and pressure of outer
disks, cooling rates are moderate, and relatively low levels of
star formation are needed to produce enough UV that heating 
balances cooling.

For outer disks where the stars and dark matter dominate gravity,
the vertical oscillation time is $t_{\rm osc}=\pi^{1/2}/(G\rho_{\rm
  sd})^{1/2}$; a dense cloud settles to the midplane in 
$\sim t_{\rm osc}/4$.  In this regime, the conversion time from gas to stars,
$t_{\rm con}\equiv \Sigma/\sigsfr$, is given by 
\begin{equation}
t_{\rm con}= t_{\rm osc}\frac{\sigma_z P_{\rm th, 0} }
{(2 \pi)^{1/2} \langle v_{\rm th}^2\rangle  \Sigma_{\rm SFR,0}},
\end{equation}
where $\langle v_{\rm th}^2 \rangle \equiv \tilde f_w c_w^2\approx c_w^2
M_{\rm diff, warm}/M_{\rm diff, total}$ is the mean thermal dispersion
in the diffuse medium (here $c_w\sim 8 \kms$ is the thermal speed in
the warm ISM).  Using $P_{\rm th, 0}\sim \langle v_{\rm th}^2 \rangle
P_{\rm gas, 0}/\sigma_z^2$ and defining a star formation energy
conversion efficiency $\varepsilon_{\rm rad} \equiv 4 \pi J_{\rm rad,
0}/(c^2 \Sigma_{\rm SFR, 0})$ for $P_{\rm rad, 0} = 4 \pi J_{\rm rad,
0}/(3c)$,
\begin{equation}
t_{\rm con}= t_{\rm osc} \frac{c }{3(2 \pi)^{1/2} \sigma_z }
\frac{P_{\rm gas, 0} }{P_{\rm rad, 0}}
\varepsilon_{\rm rad}.
\end{equation}
That is, the gas conversion time (or depletion time) is set by the
time for gas to settle to the midplane, scaled by factors for the 
ratio of gas-to-radiation pressure in the Solar neighborhood, the 
mass-to-energy conversion efficiency, and $c/\sigma_z$.

In inner disks, unlike outer disks, we have $ \sigdiff \ll \siggbc
\approx \Sigma$, so that $\sigsfr \propto \Sigma$.  In inner disks, it
is straightforward to show that there is an upper limit on the diffuse
gas surface density $\sigdiff$.  Physically, the reason for this limit
is that the diffuse-gas cooling rate per particle increases with higher
density and pressure in the inner parts of disks at least as 
$n \Lambda\propto
\sigdiff \siggbc$ (since $n \Lambda \propto \sigdiff/H \propto
\sigdiff g_z /\sigma_z^2 \propto \sigdiff \siggbc [1 + g_{\rm
  sd}/g_{\rm GBC}]/\sigma_z^2$), whereas the heating rate per particle 
varies as $\Gamma \propto \sigsfr \propto \siggbc$.  Thus, cooling will exceed
heating (causing mass to drop out of the diffuse component) unless
$\sigdiff$ is sufficiently low.  Enhanced cooling and mass ``dropout''
is likely responsible at least in part for the ``saturation'' of HI
surface densities at $\simlt 10 \Msun \pc^{-2}$ that has been observed
in the inner parts of galaxies \citep{Bigiel08}.

Based on the relations described above, the star formation law is
expected to steepen from $\sigsfr \propto \Sigma$ in inner disks 
to $\sigsfr \propto
\Sigma \sqrt{\rho_{\rm sd}}$ in outer disks.  A
reduction of the specific star formation rate $\sigsfr/\Sigma$ is
indeed observed in galaxies starting at $\Sigma \simlt 10 \Msun
\pc^{-2}$ \citep{Bigiel08,Ler08}. For some galaxies, a further
power-law relation $\rho_{\rm sd} \propto \Sigma^{2p}$ may hold such
that $\sigsfr \propto \Sigma^{1+p}$ in outer disks, but this need not
be the case in general -- that is, integrated 
``Schmidt''-type relations need not hold.

In OML, the full solution for $\sigsfr$ is obtained
as a function of $\Sigma$, $\rho_{\rm sd}$, and the parameters $\alpha
\equiv P_{\rm tot}/\Pth$ and $\tilde f_w \equiv \langle v_{\rm th}^2
\rangle/c_w^2$, under the assumptions of thermal and dynamical equilibrium
described above.  It is also shown that the theoretical solution for
$\sigsfr$ agrees well overall with a sample of disk galaxies analyzed
in \citet{Ler08}, with especially close correspondence for the large
flocculent galaxies NGC 7331 and NGC 5055.  Figure 1 shows an example of 
the comparison between the model and data, for NGC 5055.
\begin{figure}
\includegraphics[width=\hsize]{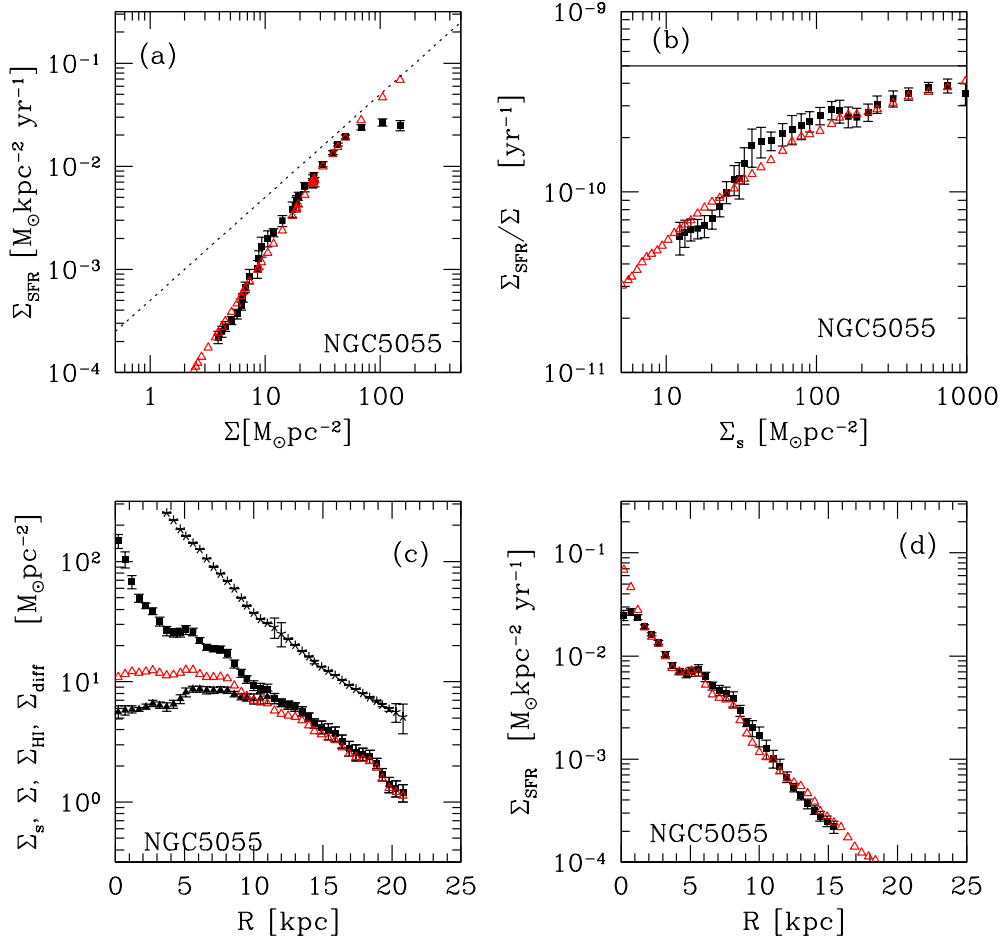}
\caption{Comparison between annular averages of the data ({\it squares}) 
for NGC 5055
\citep{Ler08}, and the thermal/dynamical equilibrium model ({\it triangles}) 
developed in OML. %\citet{2010ApJ...OML}.  
Both the star formation rates 
as a function of radius in the galaxy (panel d), and star formation
rates as a function of gas and stellar density (panels a and b) agree
with the model predictions.
}
\end{figure}

Given the promising comparisons between the analytic theory and
observations, it will be quite interesting to develop numerical
simulations that fully test the assumptions and results of the
thermal/dynamical equilibrium model.  Encouragingly, the poster
presented by C.-G. Kim at this meeting shows that initial numerical
tests support the assumptions of thermal and dynamical equilibrium
adopted in the analysis of OML.
As discussed above, the OML theory 
contains parameters that must be set from  
either observations or detailed simulations.  In the remainder
of this contribution, we review what is known in this regard based on
previous numerical work, and what measurements will be needed from
future modeling efforts.

\section{Numerical evaluation of parameters}

From equations (\ref{gz-eq}) and (\ref{ratio-eq}), the star formation rate
in outer-disk regions is expected to vary as 
$\sigsfr \propto \Sigma \sqrt{2 G \rho_{\rm sd}} \sigma_z/\alpha$, where 
$\alpha \equiv \sigma_z^2/v_{\rm th}^2$ and 
$\sigma_z^2=v_{\rm th}^2 + v_{\rm turb}^2 + (1/2)\Delta(v_A^2-2v_{A,z}^2)$,
with $v_{\rm th}^2$, $ v_{\rm turb}^2$, and $v_A^2$ the 
(mass-weighted) mean thermal, turbulent,
and Alfv\'en speeds in the diffuse gas (we now omit 
angle brackets denoting averaging).  The coefficient  
$\sigma_z/\alpha$ can also be written as 
$v_{\rm th}^2/\sigma_z = c_w^2 \tilde f_w /\sigma_z$. 
Thus, the star formation rate is expected to depend on the total
velocity dispersion $\sigma_z$ (or the ratio $\sigma_z/c_w$, where $c_w$ is 
fixed by atomic physics), and on the fraction of diffuse gas in the warm
phase $\approx \tilde f_w$ 

The ratios $\sigma_z/c_w$ and $\tilde f_w \approx M_{\rm diff,
  warm}/M_{\rm diff, total}$ depend on the details of gas dynamics in
the diffuse ISM.  Important effects include warm and cold phase
exchange via thermal instability; turbulence (with the associated
shock heating and adiabatic temperature changes, as well as turbulent
mixing); conversion of diffuse gas to GBCs via midplane settling,
self-gravity, and turbulence-induced cloudlet collisions; return of
GBC gas to the diffuse phase by photodissociation and by
``mechanical'' destruction processes (including expanding HII regions,
winds, SNe, and radiation pressure).  Turbulence in the diffuse gas
can be driven by stellar energetic inputs as well as spiral shocks, the
magnetorotational instability, large-scale gravitational instabilities
in the disk, and cosmic infall.  

Numerical studies to understand the
various effects involved are very much a work in progress, but some
consensus is already beginning to emerge on a number of points:
\smallskip
\begin{itemize}

\item For a medium with a bistable cooling curve, the
  midplane thermal pressure tends to evolve, by exchange of mass
  between cold and warm components of the diffuse phase, such that the
  mean value is comparable to, or slightly below, $P_{\rm two-phase}$
  \citep{PO05,PO07}. Since out-of equilibrium effects depend on the
  heating time from shocks compared to the cooling time, the mean value of
  the thermal pressure, as well as the breadth of the pressure
  distribution, must in general 
  be affected by the scale and the amplitude of turbulence (see
%Gazol
\citealt{2005ApJ...630..911G,2009ApJ...693..656G,
%Audit+Hennebelle
2005A&A...433....1A,
2010A&A...511A..76A,
2007A&A...465..431H,
%Joung et al
2006ApJ...653.1266J,
2009ApJ...704..137J}).  Realistic numerical evaluations of the mean 
thermal pressure (for a given radiative heating rate) therefore
will require numerical simulations in which the vertical box size is comparable
to the true scale height of the diffuse ISM, and in which the turbulent 
amplitude is $\sim 5-10\kms$.

\item Magnetic fields in differentially-rotating multiphase disks are 
amplified by the magnetorotational instability until the magnetic pressure
becomes comparable to the thermal gas pressure, with $B_z^2 \ll B^2$ 
\citep{%PO
PO05,PO07,
%Wang&Abel
2009ApJ...696...96W}.  Supernova-driven turbulence also contributes to 
amplifying the magnetic field 
\citep{
%Avillez &Breitschwerdt
2005A&A...436..585D,
%MacLow et al
2005ApJ...626..864M}.

\item The energy input from supernovae yield ISM velocity dispersions 
$\sim 5-10 \kms$ for models with a wide range of supernova driving rates 
and disk properties (e.g. 
\citealt{
%Kim 2004
2004JKAS...37..237K,
%Avillez &Breitschwerdt
2005A&A...436..585D,
%Dib
2006ApJ...638..797D,
%Shetty&Ostriker 2008
2008ApJ...684..978S,
%Agertz
2009MNRAS.392..294A,
%Joung Mac Bry
2009ApJ...704..137J}).  These values are comparable to those observed in 
the HI gas.  Simulations have also shown that the turbulent 
amplitudes decrease at smaller scales and for higher densities.  With this
range of turbulent velocity dispersions, 
the turbulent pressure in simulations of the diffuse ISM is comparable to 
the thermal pressure.

\item The interaction between self-gravity and rotational shear also drives
turbulence at significant levels ($\simgt 10 \kms$) in galactic disks 
\citep{KO01,
KO07,
%Wada Meurer & Norman
2002ApJ...577..197W,
%Shetty + O
2008ApJ...684..978S,
%Tasker&Tan
2009ApJ...700..358T,
%Agertz
2009MNRAS.392..294A,
%Aumer
2010ApJ...719.1230A,
%Bournaud
2010arXiv1007.2566B}.  However, the turbulent power is much larger at the large
($\sim \kpc$) scales that dominate the swing amplifier than at
scales below the disk scale height, and in-plane velocities (which do not 
contribute to vertical support of the disk) are much larger than vertical
velocities.  Thus, turbulence driven by instabilities on large scales is 
likely of limited importance in regulating the effective midplane
pressure (for a given local gas surface density $\Sigma$), 
and hence the star formation rate.  (Gravitational instabilities
would, however, enhance $\Sigma$ and thus $\sigsfr$ locally.) 
Flapping associated with non-steady spiral
shocks also drives turbulence in the diffuse ISM  
\citep{2006ApJ...649L..13K,2010arXiv1006.4691K}, but again, vertical motions
are small compared to horizontal motions.

\end{itemize}

Although numerical results have shown that the total turbulent
velocity dispersion $\sigma_z$ is relatively insensitive to the disk
properties and the supernova driving rate (consistent with
observations), it is much less certain how the warm fraction, or
$v_{\rm th}^2=\tilde f_w c_w^2\approx c_w^2 M_{\rm diff, warm}/M_{\rm
  diff, total}$, depends on disk conditions and/or the star formation
rate.  Assessing this dependence, including a full exploration of
parameter space, is an important task for future numerical studies.
The fraction of diffuse atomic gas in different phases is not well known
empirically, either, although observations of C II with {\it Herschel}
potentially afford a means to separate cold and warm components of the
atomic medium (which both contribute to 21 cm emission).

Finally, it remains important to understand more fully how spiral
structure develops, and in particular, whether it is possible to
characterize in a simple way the fraction of gas in a given annulus
that is found in ``arm'' vs. ``interarm'' conditions, and what the
compression factor is for the gas surface density.  Numerical
simulations have begun to marry spiral structure with 
an increasingly realistic treatment of the ISM (including multiple
phases, turbulence, and magnetic fields); much more, however, remains to 
be done on this front. It also remains to be determined
how well models like that of OML apply 
locally for galaxies with strong spiral structure.  More
generally, it is important to assess which equilibria (thermal,
dynamical, star formation) still apply locally even in galaxies with 
large-scale transient structure in the ISM (due to spiral arms, tidal
interactions, mergers, cosmic inflows, etc.).

\section{Conclusion}

Gas is the raw material for star formation, but the detailed state of
the ISM, which depends in turn on the internal 
galactic environment, determines the
rate at which this material is processed to create new stars.  Recent
observations have begun to explore the correlation between gas content
and star formation at increasingly high spatial resolution, revealing
changes in star formation ``laws'' between inner and outer disks;
other environmental dependences of star formation have also been
explored, including intriguing correlations between molecular  
and stellar content of galactic disks.  

Although the simplest recipes for star formation (such as a rate that
depends inversely on the free-fall time at the mean ISM density) have
difficulty matching the data, models that account for feedback and the
multiphase character of the ISM are more successful.  In particular,
recent work suggests that the empirical correlation between molecular
content and estimated midplane pressure can be understood as
reflecting a state of simultaneous thermal and dynamical equilibrium
in the diffuse ISM.  The thermal/dynamical equilibrium model of
OML develops the idea that UV from OB stars provides
a feedback loop that regulates the star formation rate: the
proportions of diffuse and self-gravitating gas adjust themselves so that
the heating rate (proportional to the mass of self-gravitating gas)
matches the cooling rate (proportional to the mass of diffuse gas and
to the vertical gravitational field).  The model formulated in OML
is promising in terms of explaining star-forming
behavior in observed systems.  With numerical simulations, it will be
possible to appraise -- and potentially revise -- the simplifying
assumptions and parameterizations adopted by this equilibrium model.
Time-dependent simulations will also lead to a much clearer
understanding of how GBCs form and disperse, and how their properties
and the formation/destruction timescales relate to galactic
environment.  This will aid in defining limits for applying 
equilibrium relations, while also pointing the way towards 
non-equilibrium theories of star formation.

\bigskip
{\bf Acknowledgements}: 
This work was supported by grant
AST-0908185 from the National Science Foundation, and by a fellowship
from the John Simon Guggenheim Foundation.  The author thanks the
referee for a helpful report.

\end{document}